\documentclass[manuscript,screen,nonacm]{acmart}

\usepackage{booktabs}
\usepackage{amsmath}
\usepackage{graphicx}
\usepackage{microtype}

\renewcommand\footnotetextcopyrightpermission[1]{}
\setcopyright{none}
\begin{document}

\title{Frozen Multimodal Embeddings for AI-Assisted Interview Assessment of Personality and Cognitive Ability}

\author{Kuo-En Hung}
\affiliation{\institution{Technology Application and Human Resource Development, National Taiwan Normal University}\country{Taiwan}}
\affiliation{\institution{HRDA.pro}\country{Taiwan}}
\orcid{0000-0003-2091-2747}

\author{Hung-Yue Suen}
\authornote{Corresponding author: Hung-Yue Suen, collin.suen@ntnu.edu.tw.}
\affiliation{\institution{Technology Application and Human Resource Development, National Taiwan Normal University}\country{Taiwan}}
\orcid{0000-0002-6796-2031}

\author{Shih-Ching Yeh}
\affiliation{\institution{Computer Science and Information Engineering, National Central University}\country{Taiwan}}

\author{Hsiang-Wen Wang}
\affiliation{\institution{Institute of Photonic System, National Yang Ming Chiao Tung University}\country{Taiwan}}

\begin{abstract}
Predicting psychological traits from asynchronous video interviews (AVIs) is a challenging problem in AI-assisted interview assessment because labeled datasets are limited while each response contains high-dimensional visual, acoustic, and verbal signals. This paper presents our solution for the ACM Multimedia AVI Challenge 2026, which evaluates two tasks: Track~1 predicts self-reported HEXACO personality traits from personality-related interview responses, and Track~2 classifies cognitive ability levels from structured AVI responses. We treat the problem as a small-sample representation learning task. Instead of fine-tuning large pretrained models, we use frozen multimodal encoders, including CLIP for visual features, Whisper for acoustic features and transcripts, and RoBERTa, E5, and DeBERTaV3 for textual representations, followed by low-capacity downstream models. For Track~1, our trait-specific regression and late-fusion system achieves an average validation MSE of 0.2696, improving over the official baseline of 0.3334. Ablation results show a three-step improvement from a global model (0.3189), to per-trait modeling (0.2871), to per-trait late fusion (0.2696), corresponding to a 19.1\% relative MSE reduction over the official baseline. For Track~2, a compact subject-attribute baseline reaches 0.5781 accuracy, while our multimodal ensemble reaches 0.5313, both above the official baseline of 0.4062. We interpret this result as evidence of possible subject-attribute shortcuts in the validation split rather than robust cognitive inference from AVI content. Overall, our findings suggest that AVI-based psychological assessment benefits from trait-specific multimodal modeling, but cognitive ability prediction requires careful control of dataset shortcuts.
\end{abstract}


\renewcommand{\shortauthors}{}
\renewcommand{\shorttitle}{}

\makeatletter
\let\@authorsaddresses\@empty
\def\@authorsaddresses{}
\makeatother

\maketitle

\pagestyle{plain}

\section{Introduction}

Asynchronous video interviews (AVIs) have become a widely adopted tool in modern personnel selection, enabling candidates to record responses to predefined questions at their own pace~\cite{lukacik2022void}. They are also becoming a central setting for AI-assisted assessment, automated video interview scoring, and multimodal candidate evaluation. This paper presents our solution for the ACM Multimedia AVI Challenge 2026, a benchmark task designed to assess whether multimodal signals from structured interview videos can support psychological assessment. The challenge contains two complementary tracks. Track~1 predicts four self-reported HEXACO personality traits~\cite{ashton2007hexaco} --- Honesty-Humility~(H), Extraversion~(E), Agreeableness~(A), and Conscientiousness~(C) --- as continuous scores. Track~2 classifies each candidate's cognitive ability as Low, Medium, or High.

The challenge setting is difficult because it combines limited labeled data with high-dimensional multimodal observations. Each AVI response contains visual behavior, acoustic patterns, and verbal content, but the number of labeled subjects is small relative to the dimensionality of modern video, speech, and language representations. Directly fine-tuning large pretrained models therefore risks overfitting under limited supervision~\cite{vinciarelli2014survey}. A central question in this benchmark is whether pretrained multimodal representations can be used effectively without full model fine-tuning.

Our approach addresses this problem using frozen pretrained encoders with low-capacity downstream models~\cite{liao2024benchmark}. For Track~1, we emphasize trait-specific modeling because different personality dimensions are expressed through different modality combinations. This design allows the system to improve beyond a single global model and the official baseline. For Track~2, we report a diagnostic analysis showing that a compact subject-attribute classifier achieves unexpectedly strong validation performance, which we interpret as a possible benchmark confound rather than evidence of deployable cognitive screening.

The contributions of this paper are:
(1)~we demonstrate strong benchmark improvement in an international ACM Multimedia challenge setting, reducing Track~1 validation MSE from the official baseline of 0.3334 to 0.2696, a 19.1\% relative MSE reduction;
(2)~we show through ablation that this improvement is driven by trait-specific modeling and per-trait late fusion rather than a single global predictor;
(3)~we present a reproducible frozen multimodal embedding pipeline that combines CLIP, Whisper, RoBERTa, E5, and DeBERTaV3 representations with low-capacity downstream models; and
(4)~we provide a diagnostic analysis of subject-attribute shortcuts in Track~2, showing that cognitive ability validation performance can be inflated by subject-level background variables.

\section{Challenge Tasks and Dataset}

The released AVI Challenge 2026 dataset used in our experiments contains 644 subjects, split at the subject level into training ($n{=}450$), validation ($n{=}64$), and testing ($n{=}130$) sets. The training and validation labels were available for algorithm development, whereas the testing labels were held out by the organizers and used only for official challenge submission. Unless otherwise stated, all performance values reported in this paper are based on the validation split. Each subject answered two generic and four personality-related questions. The full prompt structure is important for interpreting the task because the interview questions define the situations in which visual, acoustic, and verbal cues are elicited~\cite{zhang2026avi}. 

\begin{table}[t]
\caption{Interview prompts and task usage in the AVI Challenge 2026 dataset. Track~1 uses q3--q6 for trait-specific personality regression, whereas Track~2 aggregates responses from all six questions for subject-level cognitive ability classification.}
\label{tab:prompts}
\small
\begin{tabular}{clp{0.46\linewidth}p{0.26\linewidth}}
\toprule
Question & Type & Interview prompt & Task usage \\
\midrule
q1 & Generic &
What would you consider among your greatest strengths and weaknesses as an employee? &
Track~2: cognitive ability classification \\

q2 & Generic &
How would your best friend describe you? &
Track~2: cognitive ability classification \\

q3 & Personality &
Think of situations when you made professional decisions that could affect your status or how much money you make. How do you usually behave in such situations? Why do you think that is? &
Track~1: Honesty-Humility regression; Track~2: cognitive ability classification \\

q4 & Personality &
Think of situations when you joined a new team of people. How do you usually behave when you enter a new team? Why do you think that is? &
Track~1: Extraversion regression; Track~2: cognitive ability classification \\

q5 & Personality &
Think of situations when someone annoyed you. How do you usually react in such situations? Why do you think that is? &
Track~1: Agreeableness regression; Track~2: cognitive ability classification \\

q6 & Personality &
Think of situations when your work or workspace were not very organized. How typical is that of you? Why do you think that is? &
Track~1: Conscientiousness regression; Track~2: cognitive ability classification \\
\bottomrule
\end{tabular}
\end{table}

Track~1 uses only the four personality-related prompts (q3--q6). Each prompt is aligned with one HEXACO trait, and each trait is predicted from its corresponding response. Track~2 uses all six prompts (q1--q6), including both generic and personality-related questions, because cognitive ability is evaluated as a subject-level classification task. Accordingly, Track~1 is a question-specific personality regression task, whereas Track~2 is a subject-level cognitive ability classification task based on aggregated interview responses.

\subsection{Track 1: Self-Reported Personality Regression}

Track~1 targets four HEXACO traits~\cite{ashton2007hexaco} via validated self-report inventories: Honesty-Humility~(q3), Extraversion~(q4), Agreeableness~(q5), and Conscientiousness~(q6). The official metric is average MSE across traits.

\subsection{Track 2: Cognitive Ability Classification}
Track~2 is a three-class ordinal task (Low/Medium/High) assessed via psychometric tests and classified by professional psychologists~\cite{zhang2026avi}. Cognitive ability is a strong predictor of job performance~\cite{schmidt1998validity}. Track~2 is presented as a diagnostic analysis to expose possible subject-attribute shortcuts, not as a deployment-ready cognitive screening model.

\section{Related Work}

\subsection{Personality Computing and AVI Assessment}
Automated personality recognition from video has advanced substantially since large-scale apparent personality benchmarks~\cite{jacquesjunior2022firstimpressions}. Vinciarelli and Mohammadi identify annotation-behavior misalignment as a core validity risk~\cite{vinciarelli2014survey}. Liao et al.\ establish that pretrained representations achieve stable performance under limited supervision~\cite{liao2024benchmark}. Sun et al.\ demonstrate that modalities contribute unequally across personality traits~\cite{sun2022personality}. Hickman et al.\ show AVI systems can achieve acceptable validity while highlighting cross-question generalizability concerns~\cite{hickman2022automated}. The AVI Challenge 2025 grounded assessments in Trait Activation Theory~\cite{tett2021tat} and Behaviorally Anchored Rating Scales~\cite{zhang2025avi}; AVI Challenge 2026 extends this by shifting to self-reported ground truth and adding cognitive ability estimation~\cite{zhang2026avi}.

\subsection{Theoretical Basis for AVI-Based Trait Inference}

The theoretical basis for AVI-based assessment can be framed through the Brunswikian lens model and the realistic accuracy model. In this view, psychological attributes are distal constructs that become inferable only through observable cues; prediction is possible when trait-relevant cues are available in the environment and can be detected and utilized by the assessor or model~\cite{brunswik1956perception,funder1995realistic}. In structured AVIs, visual behavior, vocal delivery, and verbal content function as cue channels. The model does not observe personality directly; it estimates trait scores from patterns of cue utilization across these channels.

Trait Activation Theory provides an additional basis for using structured interview responses to assess personality. Traits are more likely to be expressed when the situation contains cues that make trait-relevant behavior appropriate or useful~\cite{tett2021tat}. Because AVI questions are standardized and designed to elicit job-relevant responses, they provide a more controlled situation than unconstrained social media videos. For Track~1, this is particularly important because the labels are self-reported HEXACO traits rather than observations from the observer. The task is therefore closer to latent trait inference: the system attempts to recover stable trait-related tendencies from how candidates organize, phrase, and deliver their responses, rather than merely estimating how they appear to external observers. From a lens-model perspective, observer-rated or apparent personality recognition is expected to be more directly supported by visible and audible cues because the criterion itself is formed from external cue utilization. Self-reported personality prediction is more challenging because the target reflects a latent self-description that may only be partially expressed through short structured AVI responses. This distinction makes Track~1 a stronger test of whether multimodal AVI cues can recover stable trait-related signals beyond surface impressions.

For cognitive ability, the theoretical rationale differs. Cognitive ability is commonly modeled as a hierarchical construct involving general ability and broad abilities such as reasoning, comprehension, and knowledge~\cite{carroll1993human,mcgrew2009chc}. Structured AVI responses can contain indirect indicators of these abilities, including verbal reasoning, response coherence, semantic organization, and fluency. Recent AVI research also suggests that verbal behavior is especially informative for cognitive ability assessment compared with purely paraverbal or nonverbal behavior~\cite{hickman2024smart}. This motivates our use of transcript-based representations for Track~2, while the subject-attribute baseline is retained as a diagnostic check for shortcut learning rather than as a recommended assessment model.

\subsection{Frozen Multimodal Representations}
CLIP provides transferable visual representations via image-text contrastive pretraining~\cite{radford2021clip}. Whisper provides robust acoustic features and transcripts via large-scale weak supervision~\cite{radford2022whisper}. RoBERTa~\cite{liu2019roberta}, E5~\cite{wang2022e5}, and DeBERTaV3~\cite{he2023debertav3} offer complementary textual representations. Using frozen feature extractors is appropriate for small labeled datasets~\cite{liao2024benchmark,suman2022multimodal}.

\subsection{Subject-Attribute Shortcuts in Benchmarks}
AVI-based cognitive ability assessment requires careful interpretation because benchmark labels may correlate with subject-level background variables. Algorithmic hiring systems are vulnerable to shortcut learning when models exploit such correlations instead of task-relevant evidence~\cite{raghavan2020bias}. This motivates our comparison between multimodal content-based classifiers and a compact subject-attribute diagnostic baseline.

\section{Method}

\subsection{Overview}
\begin{figure}[htbp]
  \centering
  \includegraphics[width=\linewidth]{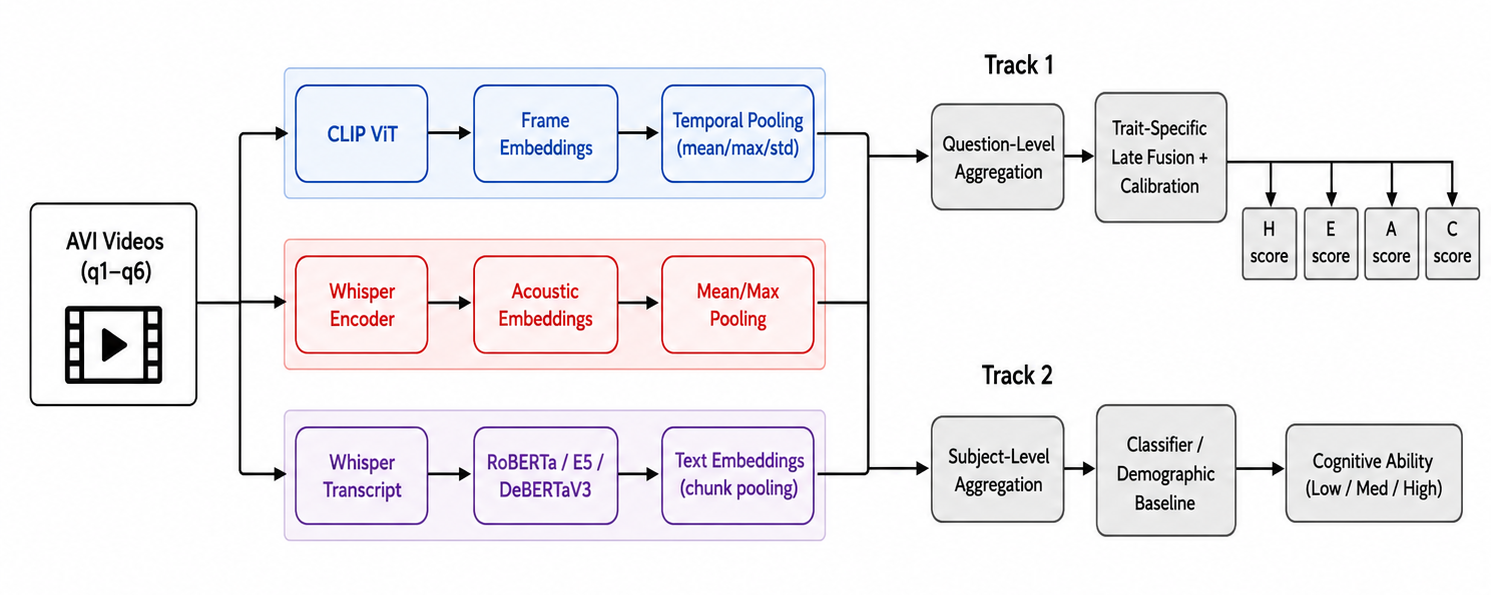}
  \Description{Architecture diagram of the proposed frozen multimodal pipeline, showing visual, acoustic, and verbal branches feeding Track 1 and Track 2 heads.}
  \caption{Overview of the frozen multimodal embedding pipeline for AVI Challenge 2026. Three parallel frozen encoder branches (visual, acoustic, and verbal) feed task-specific prediction heads for Track~1 (trait-specific regression) and Track~2 (cognitive ability classification).}
  \label{fig:pipeline}
\end{figure}

The system extracts frozen multimodal embeddings, constructs candidate feature sets, trains downstream models, and applies task-specific fusion and calibration. No pretrained encoder is fine-tuned. Figure~\ref{fig:pipeline} illustrates the overall architecture.

\subsection{Design Rationale}

The central design choice of our system is to treat the AVI Challenge as a small-sample multimodal representation problem rather than an end-to-end fine-tuning problem. Each interview response contains visual, acoustic, and verbal information, but the number of labeled subjects is limited. Under this constraint, fine-tuning high-capacity video, speech, or language models could easily overfit the validation split. We therefore keep all pretrained encoders frozen and restrict learning to low-capacity downstream models, dimensionality reduction, late fusion, and calibration.

For Track~1, we avoid using a single global personality predictor. The four HEXACO traits differ in their behavioral expression and in the modalities that are likely to carry useful signals. For example, Extraversion may be reflected in vocal and expressive behavior, whereas Conscientiousness may be more strongly associated with response structure and verbal content. The system therefore performs trait-specific model selection and fusion, allowing each trait to use a different combination of visual, acoustic, and textual evidence.

For Track~2, we explicitly compare multimodal content-based models with a compact subject-attribute baseline. This comparison is included not as a proposed deployment strategy, but as a diagnostic test for shortcut learning in the benchmark. If a small set of subject-level attributes outperforms multimodal interview content, the result suggests that validation accuracy may partly reflect split-specific correlations rather than robust cognitive inference from AVI behavior or language.

\subsection{Visual Features}
Frames are sampled at four profiles (sparse: 0.5~FPS/16~frames; lite: 0.5~FPS/32~frames; base: 1.0~FPS/60~frames; dense: 2.0~FPS/120~frames) and encoded with CLIP ViT-B/32~\cite{radford2021clip}. Embeddings are aggregated via mean, max, and std pooling plus temporal change descriptors.

\subsection{Audio and Speech Features}
Audio is divided into 30-second chunks and encoded with the Whisper base encoder~\cite{radford2022whisper}. Hidden states are pooled into acoustic embeddings capturing fluency, vocal stability, and prosodic variation. Whisper also generates transcripts for the text branch.

\subsection{Text Features}
Transcripts are encoded with three frozen language models: RoBERTa~\cite{liu2019roberta} for contextual representations, E5~\cite{wang2022e5} for dense semantic embeddings, and DeBERTaV3~\cite{he2023debertav3} for chunked long-transcript representations, avoiding truncation for longer responses. For Track~1, embeddings from the corresponding personality question (q3--q6) are used per trait; for Track~2, embeddings from all six responses are mean-pooled at the subject level.

\subsection{Track 1: Trait-Specific Regression and Fusion}
Each trait is modeled independently. Candidate regressors include Ridge~\cite{hoerl1970ridge}, PCA+Ridge, Elastic Net~\cite{zou2005elasticnet}, Bayesian Ridge, and Partial Least Squares. RidgeCV searches $\alpha \in \{0.001, \ldots, 10^6\}$; PCA dimensions from $\{32, 64, \ldots, 448\}$.

Late fusion combines validation predictions via equal-top-$k$ averaging ($k \in \{1,2,3,5\}$ selected on validation set), greedy selection, grid-searched weights, or NNLS. The fused prediction is calibrated as:
\begin{equation}
\hat{y}_{cal} = \mu_t + s \cdot (\hat{y}_{fused} - \mu_t) + b
\label{eq:cal}
\end{equation}
where $\mu_t$ is the training-set trait mean, $s \in [0.75, 2.00]$ and $b \in [-0.20, 0.20]$ are optimized via grid search on the validation set. Outputs are clipped to $[1, 5]$.

\subsection{Track 2: Classification and Diagnostic Baseline}
Two model families are evaluated: (1)~multimodal embeddings with regularized classifiers and soft-voting ensembles; (2)~compact subject-level variables (gender, age, education, work experience) with LogisticRegressionCV and PCA+LogisticRegressionCV as a subject-attribute confound diagnostic.

\section{Experimental Setup}

Track~1 model selection uses per-trait validation MSE; Track~2 uses validation accuracy. The released training set ($n{=}450$) was used for model fitting, and the validation set ($n{=}64$) was used for model selection, fusion-weight tuning, calibration, and diagnostic evaluation. The test set ($n{=}130$) was not used for model selection, calibration, or error analysis because test labels were not released to participants.
We additionally report a grouped OOF robustness check (\textit{trainval\_cv\_mean})~\cite{arlot2010survey} as a stability diagnostic, not a competing performance estimate. The official baseline uses CLIP+Whisper+RoBERTa with temporal average pooling and deep ensemble regression~\cite{zhang2026avi}, achieving Track~1 avg.\ MSE\,=\,0.3334 and Track~2 accuracy\,=\,0.4062.

\section{Results}

\subsection{Track 1: Ablation and Main Finding}

The main Track~1 result is that a frozen multimodal pipeline can substantially outperform the official challenge baseline when the downstream model is designed around trait-specific prediction. Table~\ref{tab:ablation} shows a consistent three-step improvement. A single global model already improves over the official baseline, reducing average MSE from 0.3334 to 0.3189. Modeling each personality trait separately further reduces MSE to 0.2871. Adding per-trait late fusion gives the final validation MSE of 0.2696, corresponding to a 19.1\% relative reduction over the official baseline. This pattern suggests that the performance gain comes from task-aligned modeling choices rather than from a single feature extractor or isolated classifier.

\begin{table}[t]
\caption{Ablation: effect of trait-specific design on Track~1 avg.\ MSE. Negative $\Delta$ values indicate improvement over the official baseline.}
\label{tab:ablation}
\begin{tabular}{lcc}
\toprule
System & Avg.\ MSE & $\Delta$ \\
\midrule
Official baseline~\cite{zhang2026avi} & 0.3334 & 0.0000 \\
Single global model & 0.3189 & $-$0.0145 \\
Per-trait, no fusion & 0.2871 & $-$0.0463 \\
Per-trait late fusion (final) & \textbf{0.2696} & $-$0.0638 \\
\bottomrule
\end{tabular}
\end{table}

\subsection{Track 1: Per-Trait Results}

\begin{table}[t]
\caption{Track~1 per-trait validation MSE.}
\label{tab:pertrait}
\begin{tabular}{lll}
\toprule
Trait & MSE & Main signals \\
\midrule
Honesty-Humility & 0.1921 & RoBERTa/E5/CLIP \\
Extraversion     & 0.3757 & RoBERTa/Whisper/CLIP \\
Agreeableness    & 0.3180 & RoBERTa/CLIP/DeBERTaV3 \\
Conscientiousness & 0.1926 & RoBERTa/Whisper/CLIP \\
\bottomrule
\end{tabular}
\end{table}

H and C are predicted more accurately than E and A, suggesting frozen representations capture more stable signals for these traits, while E and A may require finer social or dialogic modeling~\cite{jacquesjunior2022firstimpressions}.

\subsection{Track 1: Grouped OOF Robustness Check}

\begin{table}[t]
\caption{Track~1 grouped OOF check. Vis=visual, Aud=audio, Txt=text, WF=weighted fusion. Avg.\ CV MSE\,=\,0.3426.}
\label{tab:oof}
\begin{tabular}{llll}
\toprule
Trait & Route & Model & CV MSE \\
\midrule
H & Vis+Txt & WF & 0.3516 \\
E & Aud+Txt & WF & 0.3433 \\
A & Vis+Txt & WF & 0.4028 \\
C & Txt     & ExtraTrees~\cite{geurts2006trees} & 0.2727 \\
\bottomrule
\end{tabular}
\end{table}

Three traits show consistent multimodal fusion preference; C favors text-only with ExtraTrees~\cite{geurts2006trees}. The higher CV MSE (0.3426 vs.\ 0.2696) is expected as this protocol omits validation-supervised calibration, confirming that modality preferences are stable across splits~\cite{arlot2010survey}.

\subsection{Track 2: Diagnostic Results}

\begin{table}[t]
\caption{Track~2 diagnostic validation results.}
\label{tab:track2}
\begin{tabular}{lccc}
\toprule
System & Acc & Mac-F1 & Wt-F1 \\
\midrule
Official baseline~\cite{zhang2026avi} & 0.4062 & --- & --- \\
Multimodal ensemble & 0.5313 & 0.5208 & 0.5313 \\
Subject-Attr. LogRegCV & \textbf{0.5781} & 0.5352 & 0.5613 \\
\bottomrule
\end{tabular}
\end{table}

Both systems improve over the official baseline. However, the subject-attribute baseline outperforms the multimodal ensemble. This should not be interpreted as evidence that subject attributes are valid cognitive predictors~\cite{hickman2024smart}. A more plausible interpretation is that the validation split contains subject-attribute correlations with cognitive labels~\cite{raghavan2020bias}. We therefore treat Track~2 performance as a diagnostic result rather than evidence of robust cognitive ability inference from AVI content.

\section{Discussion}

The Track~1 results show that the key performance gain comes from task-specific downstream modeling rather than from a single pretrained representation alone. The official baseline already uses strong pretrained multimodal components, but our ablation demonstrates that model organization matters substantially. Moving from a single global model to per-trait modeling reduces average validation MSE from 0.3189 to 0.2871, and adding per-trait late fusion further reduces MSE to 0.2696. This indicates that self-reported personality prediction in AVIs should not be treated as a homogeneous regression problem. Different traits require different combinations of visual, acoustic, and verbal evidence.

The per-trait results provide further support for this interpretation. Honesty-Humility and Conscientiousness are predicted more accurately than Extraversion and Agreeableness, suggesting that the available frozen representations capture more stable signals for some traits than others. The selected feature sets also differ by trait. Honesty-Humility benefits from RoBERTa, E5, and CLIP representations; Extraversion uses RoBERTa, Whisper, and CLIP; Agreeableness relies on RoBERTa, CLIP, and DeBERTaV3; and Conscientiousness uses RoBERTa, Whisper, and CLIP. These differences suggest that trait-specific fusion is not only a performance trick but also a more appropriate modeling assumption for AVI-based personality assessment.

The use of self-reported HEXACO scores is also important to interpret these findings. Many earlier personality computing benchmarks focus on apparent personality, where models predict how observers perceive a person from video. Track~1 of the AVI Challenge 2026 instead targets self-reported personality scores, making the task closer to latent trait inference than impression prediction. This distinction may explain why textual and semantic representations remain important across traits: the model is not simply predicting visible social impressions, but attempting to recover stable trait-related signals from what candidates say, how they say it, and how they behave during structured responses.

The Track~2 results require a more cautious interpretation. Although the multimodal ensemble improves over the official baseline, the compact subject-attribute baseline performs better. This does not imply that subject attributes are valid cognitive ability predictors. A more plausible explanation is that the validation split contains correlations between cognitive labels and subject-level background variables such as age, education, or work experience. We therefore treat Track~2 as a diagnostic result. The result is useful because it reveals a potential weakness in benchmark interpretation: a model can achieve higher validation accuracy without necessarily learning cognitive ability from AVI content.

The study demonstrates that frozen multimodal representations can support competitive performance in an international ACM Multimedia benchmark when paired with careful downstream modeling. The main technical contribution is the combination of frozen visual, acoustic, and textual encoders with trait-specific regression, late fusion, calibration, and shortcut diagnostics. This design is practical for small-sample AVI assessment settings because it avoids full model fine-tuning while still allowing different psychological constructs to rely on different evidence sources.

\section{Implications}

Beyond benchmark improvement, the proposed system demonstrates a practical modeling strategy for AI-assisted video interview assessment under realistic data constraints. In many applied assessment settings, labeled interview data are limited, while each candidate response contains rich multimodal information. The frozen-encoder design addresses this constraint by reusing large pretrained visual, speech, and language models without fine-tuning them on a small challenge dataset. This makes the system more computationally efficient and reduces the risk of overfitting compared with end-to-end training.

The trait-specific fusion results also have practical implications for the design of interview assessment systems. Personality traits should not be treated as interchangeable targets for prediction. Different constructs may require different evidence sources, such as verbal semantics, acoustic fluency, visual behavior, or their combinations. This finding supports the development of modular AI interview assessment systems in which each psychological construct is modeled with task-specific feature selection, fusion, and calibration.

Finally, the Track~2 diagnostic analysis highlights the importance of validity checks in AI-assisted assessment. A model may achieve strong validation accuracy by exploiting subject-level shortcuts rather than learning task-relevant cognitive signals from interview content. For applied systems, this suggests that benchmark performance should be accompanied by shortcut diagnostics, balanced validation protocols, and careful separation between predictive accuracy and construct-valid inference.

\section{Limitations}
First, calibration parameters in Eq.~\ref{eq:cal} are selected using the validation split ($n{=}64$); therefore, the reported MSE of 0.2696 may be optimistic for unseen test data~\cite{arlot2010survey}. Second, frozen encoders and frame-level visual embeddings may miss task-specific cues such as gaze, action units, and gesture dynamics that carry predictive validity for interview assessment~\cite{liao2024benchmark,naim2018interview}. Third, Track~2 requires cautious interpretation due to subject-attribute confounding~\cite{raghavan2020bias}, and transcript features depend on ASR quality~\cite{radford2022whisper}. 

\section{Conclusion}

We presented a frozen multimodal embedding framework for the ACM Multimedia AVI Challenge 2026. For Track~1, the proposed system reduces average validation MSE from the official baseline of 0.3334 to 0.2696, achieving a 19.1\% relative MSE reduction. Ablation results show that this improvement is driven by trait-specific modeling and per-trait late fusion, indicating that self-reported personality prediction in AVIs should not be treated as a single homogeneous regression problem. Different traits benefit from different combinations of visual, acoustic, and verbal representations.

For Track~2, both the multimodal ensemble and the compact subject-attribute baseline improve over the official baseline, but the stronger performance of the subject-attribute model reveals a possible shortcut in the validation split. We therefore treat Track~2 as a diagnostic result rather than a claim of robust cognitive ability inference from AVI content.

Accordingly, these findings demonstrate our team's ability to develop a competitive multimodal AI interview assessment system under international ACM Multimedia benchmark constraints, while also highlighting the need for trait-specific modeling and careful validation when psychological labels may correlate with subject-level background variables.

\begin{acks}
The authors thank the AVI Challenge 2026 organizers for providing the dataset and baseline system. We also thank the R\&D team of HRDA.pro for supporting this work as part of HRDA's ongoing development of AI-assisted interview assessment technologies. No additional private or in-house datasets were used.
\end{acks}

\bibliographystyle{ACM-Reference-Format}
\bibliography{references}

\end{document}